\documentclass{pasj00}

\begin{document}
\SetRunningHead{Ishiyama {\it et al.}}{}
\Received{2009 August 25}
\Accepted{2009 October 1}

\title{
GreeM : Massively Parallel TreePM Code for Large Cosmological {\it N}-body Simulations
}


\author{Tomoaki \textsc{Ishiyama}, \altaffilmark{1,2}
Toshiyuki \textsc{Fukushige}, \altaffilmark{3}
and
Junichiro \textsc{Makino}, \altaffilmark{1}
}

\altaffiltext{1}{National Astronomical Observatory, Mitaka, Tokyo 181-8588, Japan}
\altaffiltext{2}{Department of General System Studies, College of Arts and Sciences,\\
University of Tokyo, Tokyo 153-8902, Japan}
\altaffiltext{3}{K\&F Computing Research Co., Chofu, Tokyo 182-0026, Japan}
\email{ishiyama@cfca.jp, fukushig@kfcr.jp, makino@cfca.jp}



\maketitle

\begin{abstract}

In this paper, we describe the implementation and performance of
GreeM, a massively parallel TreePM code for large-scale cosmological
{\it N}-body simulations.  GreeM uses a recursive multi-section
algorithm for domain decomposition. 
The size of the domains are adjusted so that 
the total calculation time of the force becomes the same for all processes.
The loss of performance due to non-optimal load balancing is around 4\%,
even for more than $10^3$ CPU cores.
GreeM runs efficiently on PC clusters and massively-parallel computers, such as a Cray XT4. 
The measured calculation speed on Cray XT4 is $5\times 10^4$ particles
per second per CPU core, for the case of an opening angle of $\theta=0.5$, if
the number of particles per CPU core is larger than $10^6$. 

\end{abstract}

\section{Introduction}
\label{sec:Introduction}

The cold dark matter (CDM) model \citep{White1978, Peacock1999} is widely 
regarded as the standard theory for the formation and evolution of the universe.  
According to this model, structure formation in 
the universe proceeds hierarchically.  
Small-scale structures form first, 
and they then merge with each other to form larger-scale structures.

Cosmological {\it N}-body simulations have been widely used to study
nonlinear structure formation in the CDM model.  A number
of numerical algorithms for cosmological {\it N}-body simulations have
been proposed.

The $\rm P^3M$ (Particle Particle Particle Mesh) algorithm, introduced by
\citet{Hockney1981}, is one such algorithm. In the $\rm P^3M$ algorithm,
gravitational interactions between particles are split into
short-range and long-range parts. The short-range part is calculated by
direct summation (PP part), and the long-range part is calculated by the PM
method (PM part), which is accelerated by Fast Fourier Transformation
(FFT).

The $\rm P^3M$ algorithm keeps the advantage of the original PM algorithm,
which uses FFT to evaluate the gravitational potential, while
improving the spatial resolution by evaluating the short-range force
directly. When the universe is close to uniform, the $\rm P^3M$ method is
very fast, since the calculation of FFT is fast and the calculation cost of PP
part is small.  However, when the system shows strong clustering, the
calculation cost for the PP part becomes very large. Thus, it is not
practical to use the $\rm P^3M$ method for a highly clustered distribution of
particles realized in CDM cosmology.

\citet{Couchman1991} developed the AP3M (adaptive $\rm P^3M$) algorithm.
In this algorithm, the gravitational interactions between particles in
high-density regions are split into three or more terms, and only
the shortest-range interaction is calculated 
directly. Intermediate-range terms are evaluated with meshes of
different sizes. Conceptually, this AP3M algorithm is simple and
efficient. In practice, efficient implementation of the AP3M algorithm on
large-scale parallel computers is not easy, and the AP3M algorithm in its
fully multi-level form is not widely used. 
GADGET-2 \citep{Springel2005} uses two-level hierarchy. 

Yet another way to reduce the calculation cost of a high-density region
of the $\rm P^3M$ scheme is to use the tree algorithm instead of direct
summation, which is now usually called TreePM \citep{Xu1995, Bode2000, 
Bagla2002, Bode2003, Dubinski2004, Springel2005, Yoshikawa2005, Khandai2009}.  
In this algorithm, the short-range interaction is calculated by the Tree
method \citep{Barnes1986}. The calculation cost per particle of the
tree algorithm is almost independent (depends only through the $\log N$
term) of the local density. Thus, the calculation cost of the PP part of
the TreePM algorithm is also nearly independent of the degree of the
clustering. 

There are two variants of the TreePM algorithm.  One is the TPM
algorithm developed by \citet{Xu1995}.  In this algorithm, the
short-range force is calculated using the tree only in high-density
regions.  TPM algorithm has been adopted by \citet{Bode2000} and
\citet{Bode2003}, and is scalable to a large number of CPU processors.

In the other variant \citep{Bagla2002, Dubinski2004, Springel2005,
  Yoshikawa2005}, the tree method is applied to the entire simulation
box, or to the domain handled by one process. The advantage of this
method over the other is the ease of implementation. The tree part
of this algorithm (which we call TreePM in this paper, while the other
we call TPM) is essentially the same as the treecode for a non-periodic
boundary condition, with just two differences. The force law is
different, and we need to handle image particles. The PM part is
relatively simple anyway. Thus, if one has already developed a
parallel treecode for a non-periodic boundary condition, it is
straightforward to implement the periodic boundary by adding the
calculation code for the PM force. In this case, domain decomposition is
based on the need of the tree part, and the PM mesh is used only for the
PM force calculation. Since parallelization of the treecode with the
non-periodic boundary condition has been very well studied, we can achieve
both the ease of implementation, and a good performance, by just using
existing algorithms.

On the other hand, in the TPM algorithm, the domain decomposition would
usually be based on the PM mesh, since there is no other data
structure to rely on. Thus, achieving high efficiency with parallel
implementation of TPM requires significant work. 

Probably for this reason, many implementations of the parallel TreePM have
been reported.  \citet{Dubinski2004} described GOTPM
(Grid-of-Oct-Trees-Particle-Mesh), which is based on one-dimensional
slab domain decomposition.  Its performance scales well on hundreds of
processes.  \citet{Springel2005} described GADGET-2 (GAlaxies with
Dark matter Gas intEracT), which uses domain decomposition based on a
space-filling curve, similar to that used by \citet{Warren1993}. 
One advantage of this method is that the
tree structure is global, and therefore the force on a particle does
not depend on the number of processes used. In some parallel
implementations of the tree algorithm, the force on a particle depends on
the number of processes used, since the way the force on a particle
is calculated is not exactly the same if the number of processes is
not the same. This makes the development and validation of the
parallel code rather troublesome.

\citet{Yoshikawa2005} presented the $\rm P^3M$ and TreePM
implementation of cosmological {\it N}-body code  on GRAPE-5
\citep{Kawai2000} and GRAPE-6A \citep{Fukushige2005} systems.  Here,
GRAPE \citep{Sugimoto1990, Makino1998, Makino2003} is used to accelerate the
tree part, in the same way as accelerating the tree algorithm for a 
non-periodic boundary condition \citep{Makino1991,Makino2004}.
We call their code YTPM. It uses one-dimensional (1-D) decomposition, 
because
it is designed for relatively small GRAPE clusters (up to 16 or 32
processors).

If the number of processes is small, 1-D decomposition is okay, simply
because other methods do not help in achieving a better parallel
performance. However, if the number of processes is large, 1-D
decomposition becomes unpractical, because of a increase in the
amount of communication and memory to store the particles and tree
nodes in boundary layers. Consider the case of a system of $N$
particles simulated on $p$ processes. The number of particles in one
process is $N/p$, and the amount of communication (and additional memory
requirement) per process per timestep is $O(N^{2/3})$ in the case of 1-D
decomposition, and $O[(N/p)^{2/3}]$ in the case of 3-D
decomposition. Thus, 3-D decomposition reduces the requirement for
communication by a factor of $O(p^{2/3})$, which can be very large,
since there are machines with more than $10^5$ processes.

We have developed a parallel TreePM code, which uses domain
decomposition based on a recursive multi-section algorithm
\citep{Makino2004}. The recursive multi-section algorithm is similar
to the widely used recursive bisection \citep{Warren1992,Dubinski1996}, 
but allows 1-D division of an arbitrary
number. Thus, it can be used on systems with the number of processes not
being powers of two. 
In addition, we modified the decomposition algorithm
so that the load balance becomes practically perfect.
Our code, which we call GreeM 
({\it G}RAPE T{\it  ree}P{\it M}), is optimized for clusters of GRAPEs or usual PC
clusters, which have a relatively poor network performance. Since the
communication performance is the limiting factor of the scalability,
our code can scale to a very large number of processes, on parallel
computers with fast networks, such as Cray XT4.

In section 2, we describe the algorithm used in GreeM.  In section 3,
the results of accuracy and performance tests are presented.  Section
4 is for summary and discussion.

\section{Algorithm}

In this section, we describe the algorithm used in GreeM.

\subsection{Gravity Force Calculation}

In GreeM, the force on a particle is divided into two components, the
PM part and the PP part. The PM part is evaluated by FFT, and the
numerical scheme is the same as that used in YTPM. The PP part
is calculated directly (actually using tree), with a modified force law
given by

\begin{eqnarray}
\label{eq:force}
{\it \bf f(r')} = \frac{m({\it \bf r}-{\it \bf r'})}{|{\it \bf r}-{\it \bf r'}|^3} g_{\rm P3M}(|{\it \bf r}-{\it \bf r'}|/\eta) ,
\end{eqnarray}
where, $\it \bf r$ and $m$ are the position and mass of a particle that
exerts a force,  $\it \bf r'$ is the position of the point at which the
force is evaluated, $ g_{\rm P3M}(R)$ is a cutoff function and
$\eta$ is the scale length for the cutoff function. The cutoff function is 
given by \citet{Hockney1981}
\begin{eqnarray}
\label{eq:plummer}
g_{\rm P3M}(R)=\left\{
\begin{array}{l}
{\displaystyle 1-\frac{1}{140}(224R^3-224R^5+70R^6} \vspace{3mm}\\
{\displaystyle + 48R^7 - 21R^8) \quad {\rm for} \quad 0 \le R < 1} \vspace{6mm}\\
{\displaystyle 1-\frac{1}{140}(12-224R^2+896R^3-840R^4+224R^5} \vspace{3mm}\\
{\displaystyle +70R^6-48R^7+7R^8) \quad {\rm for} \quad 1 \le R < 2 } \vspace{6mm}\\
{\displaystyle 0 \quad {\rm for} \quad 2 \le R} .
\end{array}
\right.
\end{eqnarray}

For calculating the PP force, we use the Phantom GRAPE library
modified for the force with cutoff (K.Nitadori et al. in preparation).
Phantom GRAPE \citep{Nitadori2006} is a highly optimized library that
calculates the gravitational interaction between particles.  It uses the 
Streaming SIMD Extension (SSE) instruction set available on recent x86
processors, which offers a much higher peak performance than the
traditional x87 instruction set.  A version of Phantom GRAPE that
uses the Altivec instruction set of the IBM POWER processor also exists.
For treecode, the effective performance of an Intel Core2 Quad (Q6600)
with four cores is comparable to that of GRAPE-6A.

\subsection{Calculation Procedure}

GreeM assumes distributed-memory parallel machines as the hardware
platform, and uses MPI as a parallel programming environment. Thus,
particles must be distributed to MPI processes. GreeM distributes
particles according to their positions, and uses recursive
multisection \citep{Makino2004} to determine the division of the simulation box.

Initially, the simulation box is divided to $p$ subboxes,
where $p$ is the number of processes.  The geometries of
subboxes are determined by an algorithm, which will be described in
subsection \ref{sec:domain}. Each subbox is assigned to one process, and each
process calculates the forces on all particles in its subbox. Thus,
processes should exchange particles so that all particles in the
subbox of one process are in the memory of that process.

After this initial decomposition is finished, we start time
integration.  The calculation for one step of time integration
proceeds in the following seven steps:

\begin{enumerate}

\item Calculation of the PM force: Each process calculates the mass
  density on the PM grid by assigning the mass of all particles using
 the  TSC (triangular shaped cloud) scheme.  For the number of PM grid
  points in one dimension, $N_{\rm PM}$, a value between $1/2H$ and
  $1/4H$ is usually used, where $H=N^{-1/3}$ is the mean distance
  between particles.   
  The scale length for the cutoff function $\eta$ and cutoff radius $r_{\rm cut}$ 
  are set to  $2\eta = r_{\rm cut} = 3/N_{\rm PM}$.
  After each process calculates the contribution
  of its particles to the PM grid, it constructs the total grid 
  by incorporating  the contributions of particles in
  other processes.
  Each process sends the values of the mass on the PM grid points to all
  other processes. 
  Each process then calculates the gravitational
  potential on the PM grid using FFT. Here, all processes perform
  exactly the same FFT calculation.  The PM forces on particles are
  calculated by interpolation  and the velocities of
  particles are updated using the PM forces.

\item Construction of the local tree: All processes construct their
  trees (local trees) from the positions and mass of their particles.

\item Exchange of the required information of global tree:
Each process sends information of its local tree required by 
other processes. This part is essentially the same as the scheme
described in \citet{Makino2004}. One process needs to receive all tree
nodes that satisfy the opening criterion from the surface of the
subbox, and are within the cutoff 
radius of the PP force, $r_{\rm cut}$, from the surface of the
subbox. Note that if one node satisfies the above criterion, it is not
necessary to send its child nodes, since that node will never be
opened in the actual force calculation. In our code, what is sent is
just the masses and positions of the tree nodes, and the tree structure
itself is not sent.

\item Reconstruction of the tree: Each process reconstructs its tree
  structure so that it contains both of its particles and tree nodes
  and particles received from other processes. We can regard this
  reconstructed tree as a ``global'' tree, since it contains 
  information of all particles in the system necessary to calculate
  the PP forces on all particles of one process.

\item Calculation of the PP force: Each process calculates the PP
  forces on its particles from the constructed ``global'' tree.  We
  use the interaction list method by \citet{Barnes1990} to improve the
  performance of GRAPE or SIMD unit of the x86 CPUs. The velocities of
  particles are updated here. We do not store the accelerations, and
  the velocities are updated twice in one timestep.

\item Position update:
The positions of particles are updated using the updated velocities.

\item Redistribution of particles:
The geometries of the subboxes are updated using new positions of
particles, and particles moved out of their original subboxes are
sent to appropriate processes.

\end{enumerate}

\subsection{Parallelization Details}

In this section, we discuss the details of parallelization that affect
the performance. First we discuss the domain decomposition, and then
optimization of the communication.

\subsubsection{Domain decomposition}
\label{sec:domain}

Our domain decomposition method is based on a recursive
multi-section algorithm \citep{Makino2004}, but we modified the basic
algorithm to improve the load balance.  The original implementation
divides the simulation box so that each subbox has the same number of
particles.  This criterion is okay for a tree algorithm with GRAPE,
since the calculation time per particle is almost independent of the
local density when we use GRAPE.  If we do not have GRAPE, even with the
tree algorithm the calculation cost depends on the local density, and
the division based on the number of particles alone is not optimal.

There have been many proposals for the method to achieve a good load
balance \citep{Warren1993, Dubinski2004, Springel2005}.  
Most of the proposed
methods are based on cost estimates based on the number of
interactions needed to obtain the forces on particles. We here
describe a much simpler, but at the same time more accurate, approach, 
which uses the measured calculation time itself as
the goal  for the load balance. 

\begin{figure*}[t]
\centering \includegraphics[height=16cm]{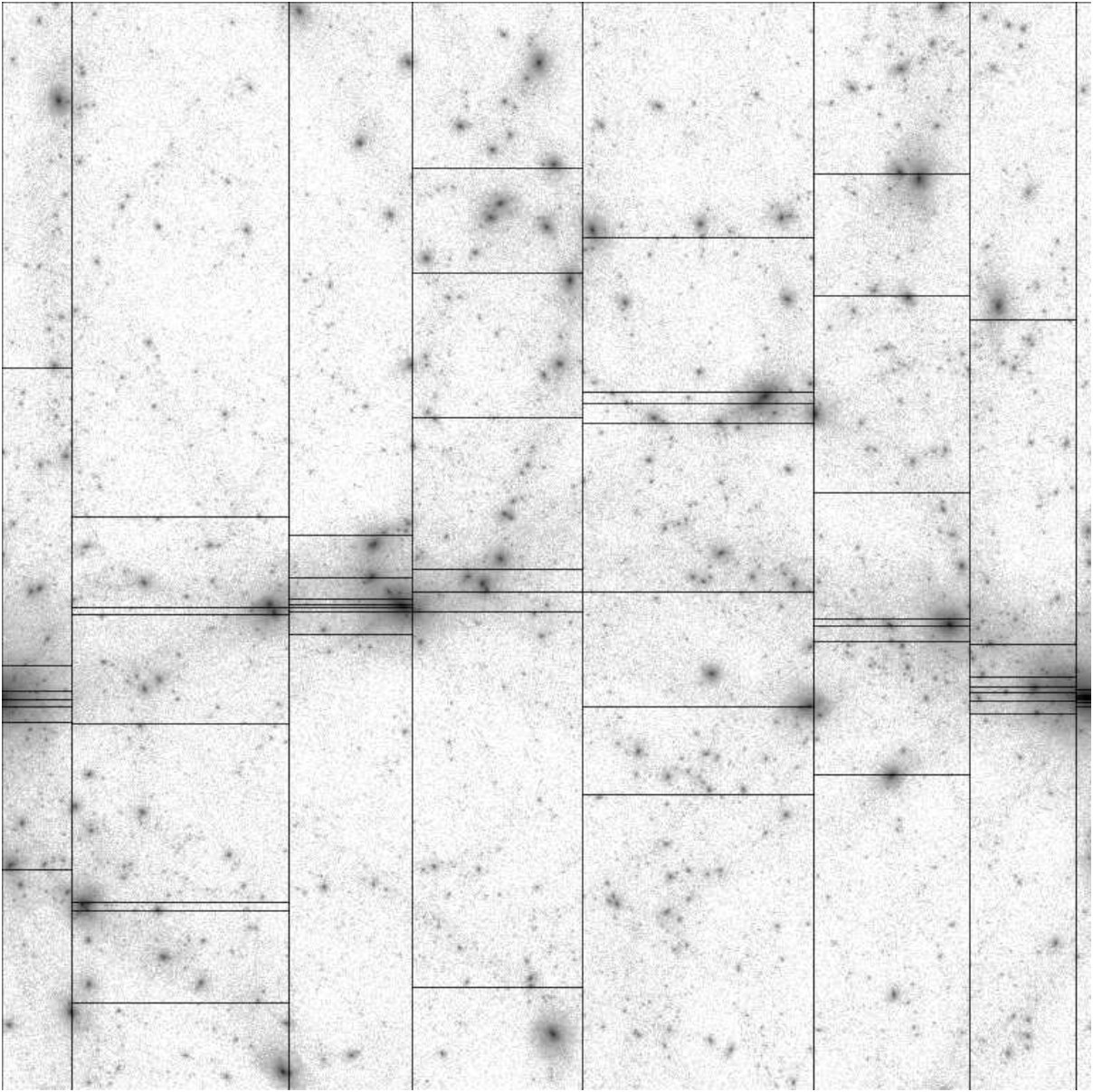}
\caption{Decomposition in the LCDM universe at $z=0$.
  It shows 8 $\times$ 8 division in two dimensions.  }
\label{pict}
\end{figure*}

In our method, we adjust the size of the domains assigned to
individual processes, so that the total calculation time of the force
(sum of the PP and PM forces) becomes the same for all processes. We
use the sampling method \citep{Blackston1997} to determine the
geometry of the domains.  In its simplest form, each process samples
particles with a fixed sampling rate, $R$, and sends them to a process.
This process then makes a division, and sends it to all other
processes.  Finally, each process exchanges particles for all other
processes according to the division.  This sampling method allows us
to drastically reduce the amount of communication that occurs for
making a division because a process does not need to know the
distribution of all particles.

\begin{table*}[t]
\begin{center}
\caption{ Maximum and minimum numbers of particles 
and calculation times. $T=t_{\rm pp} + t_{\rm pm}$
for a $1000^3$ dark matter simulation  with 512 CPU cores at $z=0$. 
}
\label{tab1}
\begin{tabular}{lccc}
\hline \hline 
Correction factor & $N_{i}/N$ & Eq. (\ref{eq:nsample_t})
& Eqs. (\ref{eq:nsample_t}) and (\ref{eq:nsamplelowlim})\\ 
\hline
$N_{\rm min}$ & 1941414 & 1433413 & 1431633  \\
$N_{\rm max}$ & 1965026 & 2548045 & 2359332 \\
\hline
$T_{\rm min}$ & 29.84 & 38.07 & 35.81\\
$T_{\rm max}$ & 50.96 & 40.39 & 41.18 \\
\hline
\end{tabular}
\end{center}
\end{table*}

A naive way to take into account the calculation cost of particles in
the sampling method would be the following. The process that makes
the domain decomposition collects both the sampled particles and their
calculation cost, and determines the geometry of the domains while taking into
account the calculation cost. This scheme would work fine, 
but it is rather
complicated.  We achieve the same effect not by assigning the cost to
the sampled particles, but by changing the sampling rate of the particles
according to the calculation cost.

The number of particles, $n_{{\rm samp},i}$, sampled on the $i$-th process
is determined as
\begin{equation}
n_{{\rm samp},i} = N R_{\rm samp} f_{{\rm samp},i},
\label{eq:nsample}
\end{equation}
where $N$ is the total number of particles, 
$R_{\rm samp}$ is the global sampling rate, and $f_{{\rm samp},i}$ is
a correction factor needed to achieve a balanced state. We chose $R_{\rm
  samp}$ so that the cost of the calculation and communication is small,
and yet $N R_{\rm samp}/p$, where $p$ is the number of processes, is
large enough that the fluctuation due to sampling is small. We typically
use $R= 4\times10^{-4}$. 

We use the
following formula to determine the correction factor:
\begin{equation}
f_{{\rm samp},i} = \frac{t_{{\rm PP},i} + t_{{\rm PM},i}}
{\displaystyle \sum_j \left( t_{{\rm PP},j} + t_{{\rm PM},j} \right) },
\label{eq:nsample_t}
\end{equation}
where $t_{{\rm pp},i}$ and $t_{{\rm pm},i}$ are the CPU time for PP 
and PM part of the $i$-th process, respectively.
By this strategy, any imbalance in the CPU time will be corrected. If
the calculation on one process takes a time longer than the average, it will
sample more particles than average. When the new domain decomposition
is created, it is adjusted so that all domains have the same number of
sampled particles. Therefore, the size of the domain for this process
becomes somewhat smaller, and the CPU time for the next timestep is
expected to become smaller.

Figure \ref{pict} shows the domain decomposition for a simulation of  a LCDM 
universe at $z=0$.  It shows 8 $\times$ 8 division in two dimensions.
The total number of particles is $256^3$.
We can see that high-density regions, such as halo centers are divided into
small boxes. In this case, the maximum number of particles in a box is
379569 and the minimum is 189901.

\begin{figure}[h]
\centering \includegraphics[height=6cm]{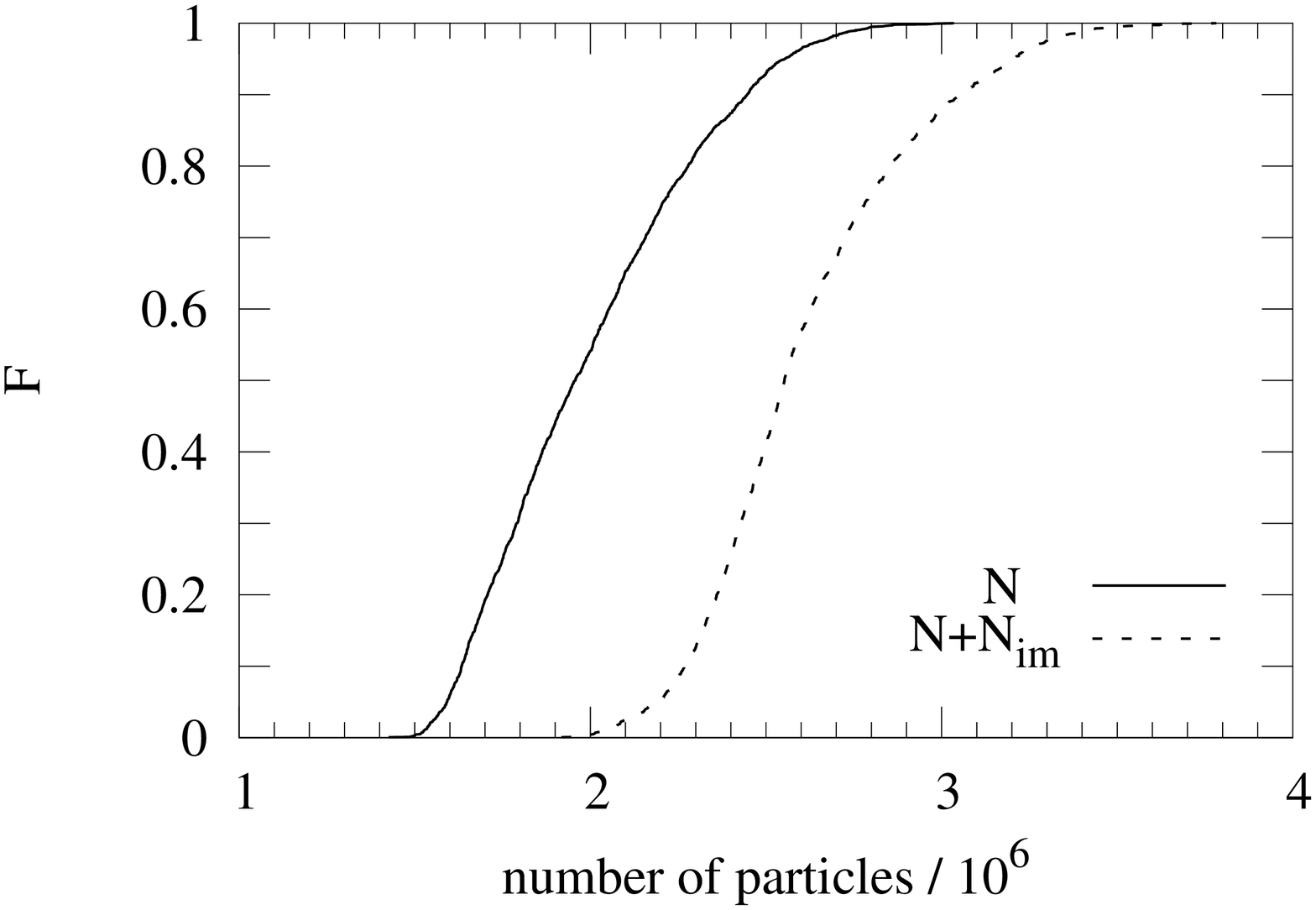}
\caption{
Cumulative fraction $F$ of CPU cores as a function of the number of particles per process.
Here, $N$ is the number of particles per process, 
and $N_{\rm im}$ is the number of particles imported from all other processes, 
respectively. 
}
\label{fig:npart}
\end{figure}

One potential problem of our method, or any method that aims to
achieving a good load balance, is the imbalance in the memory usage.
Figure \ref{fig:npart} shows the cumulative distribution of 
the number of particles per core,
for an \citet{Ishiyama2009} simulation 
($N=1600^3$, $z=0$, $\theta=0.5$, and $p=2048$) .
In this example, the process with the maximum number of particles
contains about 50\% more particles than
average (3035116 compared to 2000000). Therefore, we need 
50\%  more memory to store the particles. Since the memory
available to individual MPI processes is usually fixed, we need this 50\% more
memory for all processes, and for most of processes this additional
memory is not used.

If the amount of memory is critical, it is easy to place the upper
limit to the number of particles on one process, by placing a lower
limit on the number of  particles to be sampled given by
\begin{equation}
\underbar{\it n}_{{\rm samp},i} = \frac{N_i R_{\rm samp}}{1+\alpha},
\label{eq:nsamplelowlim}
\end{equation}
where $N_i$ is the current number of particles in process $i$, and
$\alpha$ is a parameter that controls the maximum number of particles
for one process. If $n_{{\rm samp},i}$, calculated using equation
(\ref{eq:nsample}), is smaller than $\underbar{\it n}_{{\rm samp},i}$, we
use the latter value as the number of particles to be sampled.
If we set $\alpha=0.2$, the maximum number of particles
in one process is adjusted so that it does not exceed the average
value by more than 20\%.

One might think that this limit on the number of particles for one
process would cause a significant degrade of the performance. In
practice, however, the degrade is very small. The reason is the following. When
we set the upper limit to the number of particles for one process,
particles that would be there need to be redistributed to other
processors. The average increase in the number of particles on other
processes is given by 
$f_{\rm over}[\langle N_{\rm  over} \rangle -N/p(1+\alpha)]/(1-f_{\rm over})$, 
where $f_{\rm over}$ is the
fraction of processes with the number of particles being more than the specified
limit, and $\langle N_{\rm  over} \rangle$ is the average number of particles for
these processes. We found $f_{\rm over}\sim 0.2$ and 
$\langle N_{\rm  over} \rangle \sim 1.3N/p$ 
when we chose $\alpha=0.2$, in our large cosmological
calculation. Therefore, the increase in the calculation time is less
than 2\%.

Table \ref{tab1} gives the maximum and minimum numbers of particles
and calculation times, $T=t_{\rm pp} + t_{\rm pm}$, for a $1000^3$ dark
matter simulation with 512 processes at $z=0$.  We used $\alpha=0.2$.
We measured them for three different load-balance schemes. The first
one (second column) is the simplest one which assigns the same number
of particles to all processes. The second one is based on the optimal
load-balance scheme of equation (\ref{eq:nsample_t}). The third one is a 
modified one with equation (\ref{eq:nsamplelowlim}).

We can see that the use of equation (\ref{eq:nsamplelowlim}) reduces
the maximum number of particles per process from $2.55 \times 10^6$ to
$2.36 \times 10^6$. This number is very close to $(1+\alpha)N/p =
2.34\times 10^6$. On the other hand, the increase in the calculation
time is actually less than 2\%. Thus, by combining equations
(\ref{eq:nsample_t}) and (\ref{eq:nsamplelowlim}), we can achieve
close-to-optimal use of both the memory and the CPU time.

In figure \ref{fig:npart} we also show the distribution of $N+N_{\rm
im}$, where $N_{\rm im}$ is the number of particles imported from
other processes to construct the global tree. In this case, $N_{\rm
im}$ is around one quarter of the average number of particles in one
process. However, since $N_{\rm im}$ is proportional to $N^{2/3}$,
when we use a large fraction of the memory available for one process,
the fraction of the memory used for $N_{\rm im}$ becomes small.

\subsubsection{Communication}

Inter-process communications occur in four places of the code.  
In order to achieve  a high performance,  it is crucial to reduce the
communication costs.

First communication occurs in the PM part. In our current
implementation, all processes receive 
the entire grid
from all other processors, 
and the amount of communication for one process is
$O(N_{\rm PM}^3)$. With an optimum implementation of parallel FFT,
the amount of communication for one process can be reduced to 
$O(N_{\rm PM}^3/p)$. However, we so far have not tried to use parallel
FFT, because we can reduce this part by just making $N_{\rm PM}$
small. A reduction of $N_{\rm PM}$ causes an increase in the
calculation cost of the PP part, but it is rather modest
because of use of the tree. 

Second communication is the exchange of tree information in the PP
part.
Each process imports tree structures in all other processes as
superparticles.  The amount of communication 
depends on many factors, but most importantly on the cutoff radius, 
the opening parameter for the tree, and the surface area of the domains
for processes. Thus, adaptive three-dimensional space decomposition is
critical here. 

The third one is that for the sampling method. Typically, we can make
this much smaller than the rest.

The fourth one is the redistribution of particles after the new domain
geometry is determined. In the case of cosmological $N$-body
simulations, this part is very small, because the velocity of
a particle, relative to the simulation box, is tiny in cosmological
simulations.

Thus, the communication of the tree structure is the most
expensive. As shown later, in our implementation the time for this
part is less than 2\% of the time for the PP force calculation, if the
number of particles per process is  $10^6$ or more. 

\subsection{Softening and Timesteps}

Time integration is performed in comoving coordinates.  We usually
use shared and time-dependent Plummer softening, $\varepsilon(z)$, given
by equation (\ref{eqn:softening}), which is similar to those used in
\citet{Kawai2004} and \citet{Kase2007}. It is given by
\begin{eqnarray}
\label{eq:softening}
\varepsilon(z)=\left\{
\begin{array}{r}
{\displaystyle \frac{1+z_{\rm crit}}{1+z}\varepsilon_{\rm fin}} \quad (z \ge z_{\rm crit})  \vspace{3mm}\\
{\displaystyle \varepsilon_{\rm fin} \quad (z<z_{\rm crit})} ,
\end{array}
\right. 
\label{eqn:softening}
\end{eqnarray}
where $\varepsilon_{\rm fin}$ is the softening for $z=0$.  The
softening is constant up to $z=z_{\rm crit}$ in comoving coordinates.
After $z=z_{\rm crit}$, it is constant in
physical coordinate.

The timestep $\Delta t(z)$ is also shared. It is  adaptive and
calculated by the following formula:
\begin{eqnarray}
\label{eq:timesteps}
\Delta t(z)=\zeta \min_i\left( \sqrt{\frac{\varepsilon(z)}{|{\bf a}_i|}} \thinspace,
\thinspace \frac{\varepsilon(z)}{|{\bf v}_i|} \right) ,
\end{eqnarray}
where $\zeta$ is an accuracy parameter; ${\it \bf v}_i$ and ${\it \bf a}_i$ are
the velocity and acceleration vector of  particle $i$.  
We usually set $\zeta=1.0$ or $\zeta=2.0$.

\section{Accuracy and Performance}

In this section, we present the result of measurements of the
accuracy and performance of our GreeM code.  We used LCDM
($\Omega_0=0.3$, $\lambda_0=0.7$, $h=0.7$, $\sigma_8=0.9$) models
consisting of $128^3$, $256^3$, and $512^3$ particles for measuring
the performance.  The box size was 107{\rm Mpc} for all models.  To
generate initial particle distributions, we used the GRAFIC package
\citep{Bertschinger2001}.

\subsection{Accuracy}

In this section, we discuss the accuracy of GreeM. First, we present
the numerical accuracy of the Phantom GRAPE library, and then the
overall accuracy of the force obtained with GreeM.

\subsubsection{Pairwise force error}

The Phantom GRAPE library for the force with cutoff uses
single-precision numbers to express the position data. Thus, if we
simply convert the original double-precision data to single-precision
data, the roundoff error after the subtraction ${\bf x}_j - {\bf x}_i$ 
can be rather large. 
Here, ${\bf x}_i$ is the position of the point at which the force is evaluated,
and ${\bf x}_j$ is the position of a particle that exerts the force.
In order to reduce the roundoff error, we
pass the shifted values, ${\bf x}_i - {\bf x}_g$ and ${\bf x}_j -
{\bf x}_g$, to the Phantom GRAPE library. 
Here, ${\bf x}_g$ is the position of the center of the particle groups in Barne's algorithm.
By this treatment, we can
reduce the roundoff error after subtraction by a factor
proportional to the size of the box for the group, which is on the
order of $N^{-1/3}$. Without this treatment, the roundoff error of
Phantom GRAPE can be dangerously large. 

\begin{figure}[t]
\centering 
\includegraphics[width=9cm]{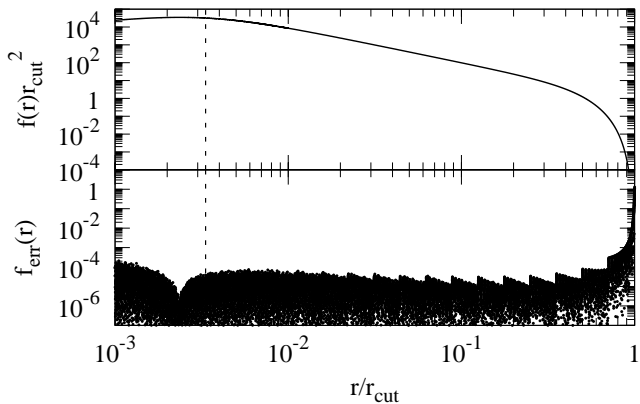}
\caption{Pairwise force error of Phantom GRAPE.  
The solid curve shows the force given in equation (\ref{eq:force}).  
The black dots show the relative pairwise force error defined in
equation (\ref{eq:accuracy}).   The vertical dashed line indicates the
value of  the force softening.
}
\label{fig2}
\end{figure}

Figure \ref{fig2} shows the error of the force between two particles
as a function of the distance. The error is defined as
\begin{eqnarray}
\label{eq:accuracy}
f_{\rm err}(r) = \frac{|f_{\rm pg}(r)- f(r)|}{|f_{\rm pg}(r)|} ,
\end{eqnarray}
where $f$ and $f_{\rm pg}$ are forces calculated in standard double
precision and that calculated with Phantom GRAPE, respectively.  We
set the cutoff radius, $r_{\rm cut}$, to $5.859375 \times 10^{-3}$ and the softening length
to $1.953125 \times 10^{-5}$. These are typical values we use in actual simulations in
which the box size is normalized to unity.

The particle distributions and calculation procedure mimic that which
appear in the tree algorithm with Barnes' vectorization algorithm.
First, we select the position of the center of the origin of the particle
groups used in Barnes' algorithm. For simplicity, we assume that the size
of the box is 1/128. This position, ${\bf x}_g$, is generated from the
uniform distribution within a cube of unit size. We then, generate the
position of one particle, ${\bf x}_i$, from the uniform distribution
within the cube of size 1/128, with the center of the box at ${\bf
  x}_g$. Next, we generate
positions of the other particles, ${\bf x}_j$, so that the position
vector relative to the first particle has a random orientation and the
logarithm of the distance follows a uniform distribution between
$10^{-6}$ and unity. We generated 256 values for ${\bf x}_i$ and 
1024 for ${\bf x}_j$.

In figure \ref{fig2} we plot the results of all pairwise force error
measurements. When the relative distance is larger than the softening
length, the typical relative error is around $10^{-5}$, but it becomes
larger for a distance close to unity. This increase is due to the fact
that the PP force approaches to zero at the distance unity. If we
measure the error relative to the pure $1/r^2$ force, there would be
no significant increase in the error. When the distance is smaller
than the softening length, the relative error is smaller, because in
this region the only source of error is rounding of the two
position vectors.
Thus,  the relative error of Phantom GRAPE library is around
$10^{-5}$ for the range of the distance that is relevant. This error
is much smaller than what is necessary in cosmological $N$-body
simulations.  

\subsubsection{Total force error}
\label{sect:totalforceerror}

In this section we discuss the distribution of the relative error of
the total force calculated with the TreePM algorithm used in GreeM.
We define the relative force error, $\Delta f_i$, of the $i$-th
particle as
\begin{eqnarray}
\label{eq:ewald}
\Delta f_i = \frac{|f_{{\rm TreePM},i}-f_{i}|}{|f_i|},
\end{eqnarray}
where $f_{{\rm TreePM},i}$ and $f_{i}$ are the acceleration of
the $i$-th particle calculated by TreePM and the exact force,
respectively. In order to estimate the exact force, 
we used a direct Ewald summation \citep{Ewald1921,Hernquist1991}. 
The scale length of the Gaussian function used in the Ewald method is 0.1, 
the real-space cutoff is 0.2, and the wavenumber cutoff is 7.14.

With Phantom GRAPE, we use the interaction list method
\citep{Barnes1990} to reduce the cost of tree traversal. We use
$N_{\rm crit}=300$ as the criterion for the grouping of
particles. With this criterion, the average number of particles that
share the same interaction list is $\approx N_{\rm crit}/4$
\citep{Makino1991}.

When we construct the tree, we stop the subdivision of a node if the
number of particles in that node is less than $N_{\rm leaf}$. By this
method, we can reduced the amount of memory required to store the tree
and to reduce the CPU time for tree construction. However, a very large
value of $N_{\rm leaf}$ causes an increase in the CPU time for the PP
force calculation. For the timing benchmarks, we used $N_{\rm
  leaf}=10$. The number of particles used here is $128^3$.

\begin{figure*}[t]
\includegraphics[height=6.6cm]{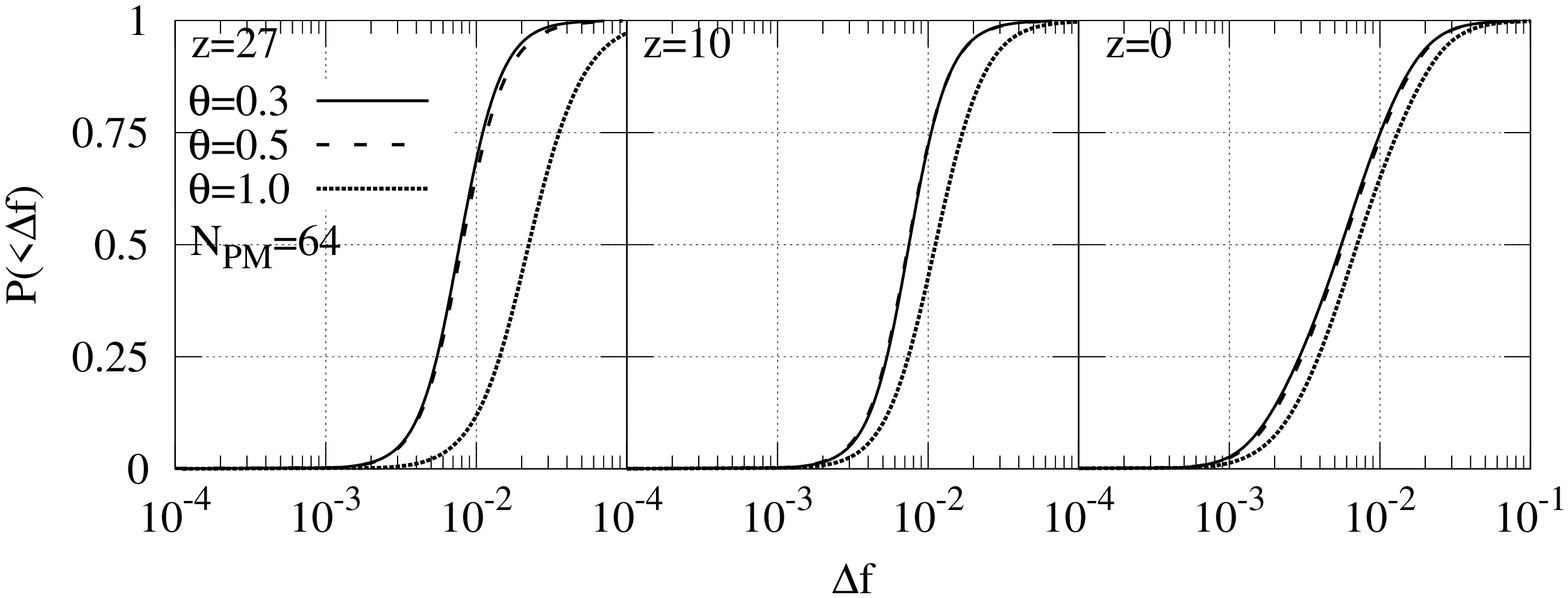}
\includegraphics[height=6.6cm]{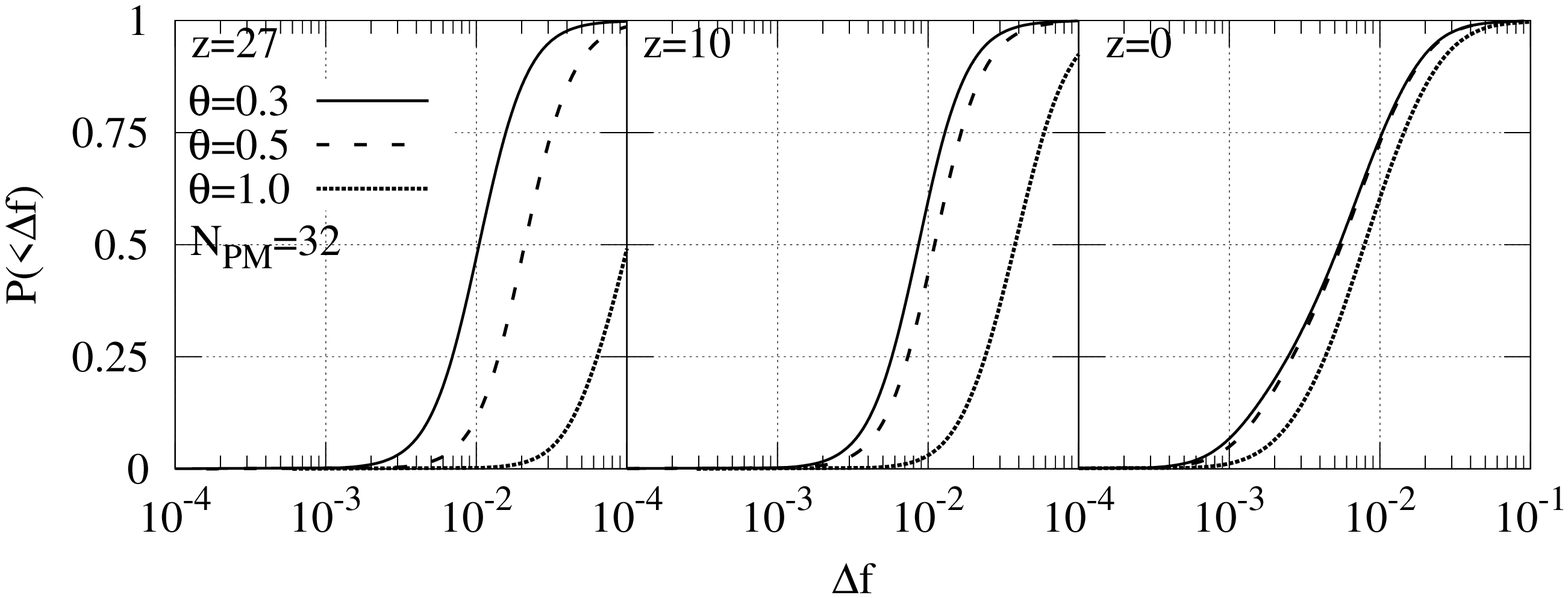}
\includegraphics[height=6.6cm]{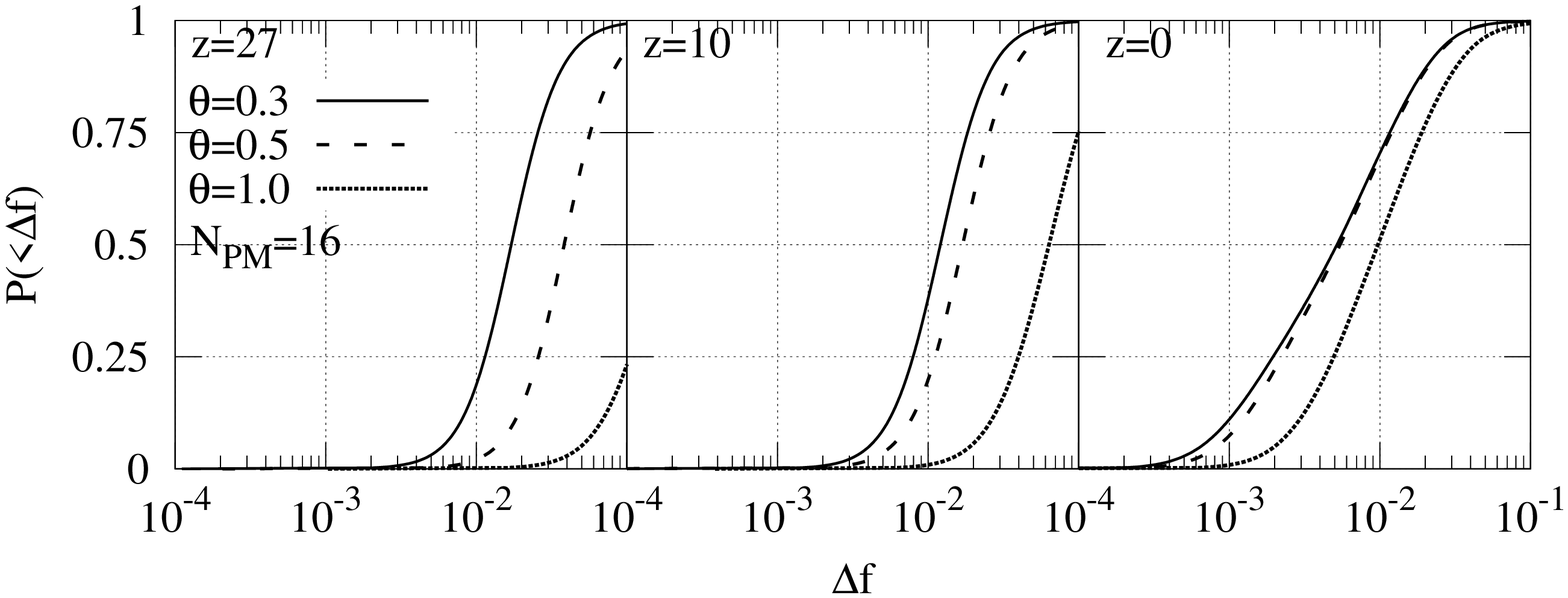}
\caption{ Cumulative distributions of the relative force errors for
  $N=128^3$ model. Top, middle, and bottom rows show the results with
  $N_{\rm PM}=64$, 32 and 16. Within each row, the left, center, and
  right panels show the result at $z=27$, 10 and 0. The solid, dashed, and
dotted curves in each panel show the result with $\theta=0.3$, 0.5 and
1.0.}
\label{fig2-2}
\end{figure*}

Figure \ref{fig2-2} shows the cumulative distribution of the relative
error. The top, middle, and bottom rows show the results with $N_{\rm
  PM}=64$, 32 and 16, and the left, middle, and right panels in one row show
the results at $z=27$, 10, and 0, respectively. In each panel, results
with $\theta=0.3$, 0.5 and 1.0 are shown.

The relationship between the accuracy of the force and the accuracy of the
result is not well understood. Here, we use the condition that the
error of 90\% of the particles is less than 2\%, which is probably too
stringent.

For $N_{\rm PM}=64$, we can achieve this goal with $\theta=0.5$, even
at $z=27$, and in the case of $N_{\rm PM}=32$ with $\theta=0.3$.  With
$N_{\rm PM}=16$, at $z=27$ we would need $\theta =0.2$.  For lower
values of $z$, we can use a significantly larger $\theta$.

These behaviors of errors are quantitatively in fair agreement with
those in previous studies \citep{Bagla2002, Wadsley2004}. For a similar
choice of parameters, the error of the Gasoline code \citep{Wadsley2004}
seems to be somewhat larger than that of ours. This could be due to the
difference in the distribution of particles. However, since they use the
Ewald method, the cutoff length of the real-space PP force is much
larger than what we used. This difference in the cutoff length might
be the cause of the difference between our result and that of Gasoline. 

Note that in the case of $N_{\rm PM}=64$, the error is almost the same
for $\theta = 0.3$ and 0.5 for all values of $z$. Also, for the case
of $z=0$, the error is almost the same for $\theta = 0.3$ and 0.5.
This means that for these cases the error is dominated by a mismatch
between the PM force and the PP force.

From the point of view of the performance, it is desirable to use a
smaller $N_{\rm PM}$, since that would reduce the memory requirement,
the amount of communication, and the calculation cost of FFT. However, as we
can see from figure \ref{fig2-2}, the error, especially at high-$z$,
increases rapidly as we reduce $N_{\rm PM}$.

\begin{figure}[h]
\centering \includegraphics[width=9cm]{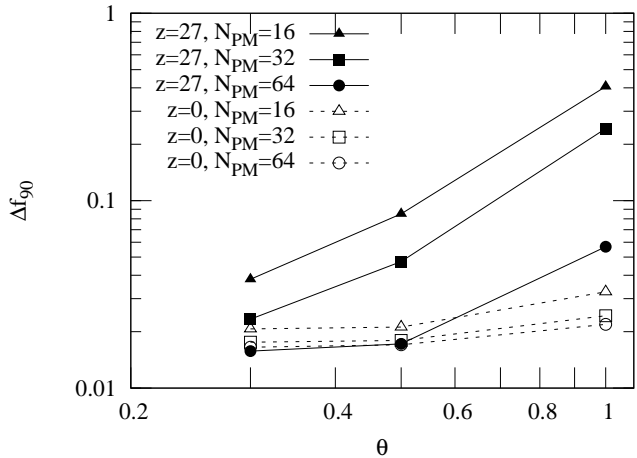}
\caption{Error of 90\% of the particles as a function of $\theta$.
The solid and dashed curves show the results at $z=27$ and $z=0$,
respectively. The triangles, squares, and circles show the results with
$N_{\rm PM}=16$, 32 and 64, respectively. 
  
}
\label{fig2-3}
\end{figure}

\begin{figure}[h]
\centering \includegraphics[width=9cm]{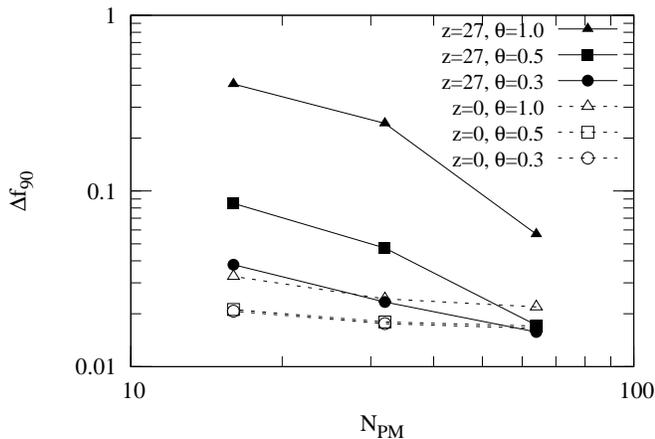}
\caption{Error of 90\% of the particles as a function of $N_{\rm PM}$.
The solid and dashed curves show the results at $z=27$ and $z=0$,
respectively. The triangles, squares, and circles show the results with
$\theta=1.0$, 0.5 and 0.3, respectively. 
}
\label{fig2-4}
\end{figure}

Figure \ref{fig2-3} and \ref{fig2-4} show the error at 90\% of the
particles as functions of $\theta$ and $N_{\rm PM}$, respectively.  We
can see that the error at $z=27$ is roughly proportional to $\theta^2$ and
$N_{\rm PM}^{-1}$. This behavior can be understood as follows.

Consider the extreme case of a purely uniform distribution of mass
without any density perturbation. In this case, the exact force is
zero, and therefore we cannot define the relative error. However, we
can still discuss the absolute error.

The error of the force from a single tree node with mass $m$ and size
$l$ at distance $r$ is dominated by the second-order (quadrupole)
term, since we use the center-of-mass approximation.  In the case of a
pure $1/r$ potential, the quadrupole term vanishes in the limit of
uniform density \citep{Barnes1989}, but in the case of the force with
cutoff the second-order term does not vanish, and this is the leading
error term.  Thus, the absolute amount of the error of the force from
one node is proportional to $mr^{-2}g_{\rm P3M}(r/\eta)(l/r)^2$.  If
we assume a uniform density of $\rho$, $m \sim \rho l^3$, and for a
given opening angle we have $l \sim r\theta$. Therefore, the error of
the force from a node at distance $r$ is proportional to $\rho r
\theta^5 g(r/\eta)$. We can see that the error is largest at
$r\sim \eta$. The number of tree nodes with $r\sim \eta$ is
proportional to $\theta^{-3}$, and we cannot assume that the errors from
different cells are random, since all cells essentially have the same
second-order terms. Therefore, the total error is proportional to
$\rho \eta \theta^2$.

We conclude that if we are to use a large PM grid spacing (more than
$4H$), the opening angle should be set to be less than 0.3 from
initial to $z=10$.  From $z=10$, we can use the opening angle around
0.5.  If we use a small PM grid spacing (less than $2H$), the opening
angle should be set to be less than 0.5 from initial to $z=10$. 

\subsection{Performance}

In this section we report on the measured performance of GreeM.  
We used a
Cray XT4 at Center for Computational Astrophysics (CfCA), National
Astronomical Observatory of Japan for the measurement.  It consists of
740 Opteron quad-core processors at a clock speed of 2.2 GHz (the
total number of cores is 2960) and 5.7TB of memory.  The peak performance is 26
Tflops.  Processors are connected in a 3D torus network with the Cray
SeaStar2 chip.  The peak bandwidth of a single link of the torus
network is about 7.6Gbyte/sec.

First, we present a result of the measurement of the parallel
performance (scaling of the performance). We then go into the details,
such as a breakdown of the CPU time, the dependence on the opening
angle, that on the distribution of particles, and memory usage.

\subsubsection{Scalability}

For measuring of the scalability, we used $256^3$, $512^3$ and
$1000^3$ particles, and $N_{\rm PM}=N^{1/3}/2$. For $1000^3$
particles, we used $N_{\rm PM}=N^{1/3}/4$, to save the memory for the
PM grid.  We measured the CPU time at $z=0$.  The parameters of the tree
parts were $\theta=0.5$, $N_{\rm crit}=300$, and $N_{\rm leaf}=10$.

\begin{figure*}[t]
\centering \includegraphics[width=12.5cm]{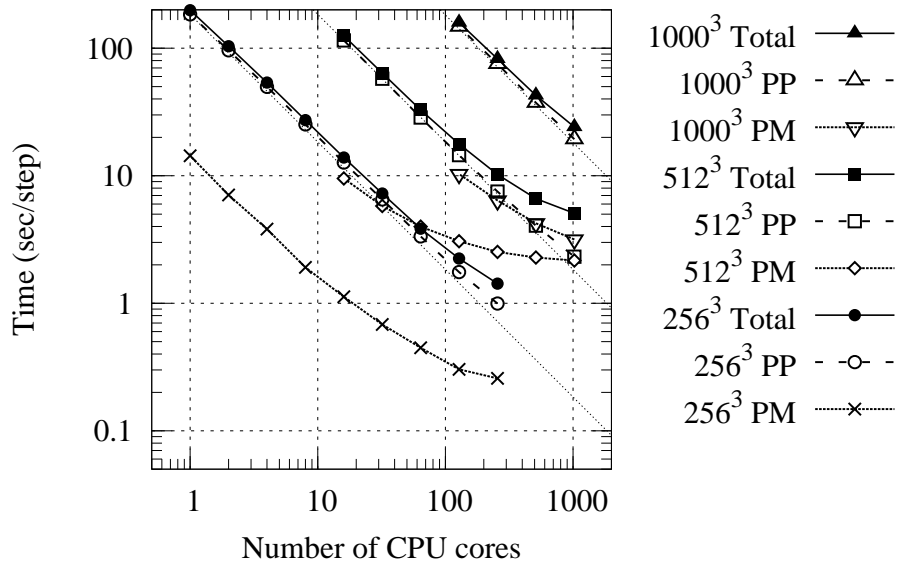}
\caption{ 
Calculation time per step of our code as a function of the
number of CPU cores on Cray-XT4.  
The curves with filled triangles, squares and circles 
show the results for total calculation time of
$1000^3$, $512^3$, and $256^3$ dark matter simulations, respectively.
The curves with open triangles, squares and circles 
show the results for the PP part, respectively.
The curves with open inverted triangles, argyles and crosses 
show the results for the PM part, respectively.
}
\label{np-t}
\end{figure*}

Figure \ref{np-t} shows the CPU time per step as a function of the
number of CPU cores.  In the case of $256^3$ particles, the parallel
speedup is almost perfect for up to 64 cores, and even with 256 cores
the gain is good.  For $512^3$ particles, parallel speedup is fine
with up to 256 cores, but degrades with more cores. With $1000^3$
particles, parallel speedup is almost perfect for the maximum number
of cores we used for the test (1024 cores).

The leveling-off of the parallel speedup comes from a leveling-off
of the CPU time of the PM part. As discussed earlier, the time for 
communication and the FFT operation of the PM part is independent of the
number of cores, since these parts are not parallelized. Thus, for
a  large number of cores, the CPU time for these parts
starts to limit the speedup. For $256^3$ and $512^3$ particles, the
ratio between the total number of particles and the total number of PM
grid points is the same. Thus, the dependence of the parallel speedup
on the number of cores is also roughly the same. In the case of
$1000^3$ particles, we reduced the number of PM grids, which is the
reason why the parallel speedup became improved. As discussed in subsection
\ref{sect:totalforceerror}, the error of the calculated gravitational
force is proportional to the inverse of the grid point. Therefore,
when we use a small PM grid, we should use a small opening angle for the
tree part to retain accuracy. This choice might result in an increase
of the total CPU time, even though the parallel speedup is improved.

\begin{figure*}[t]
\centering \includegraphics[width=12.5cm]{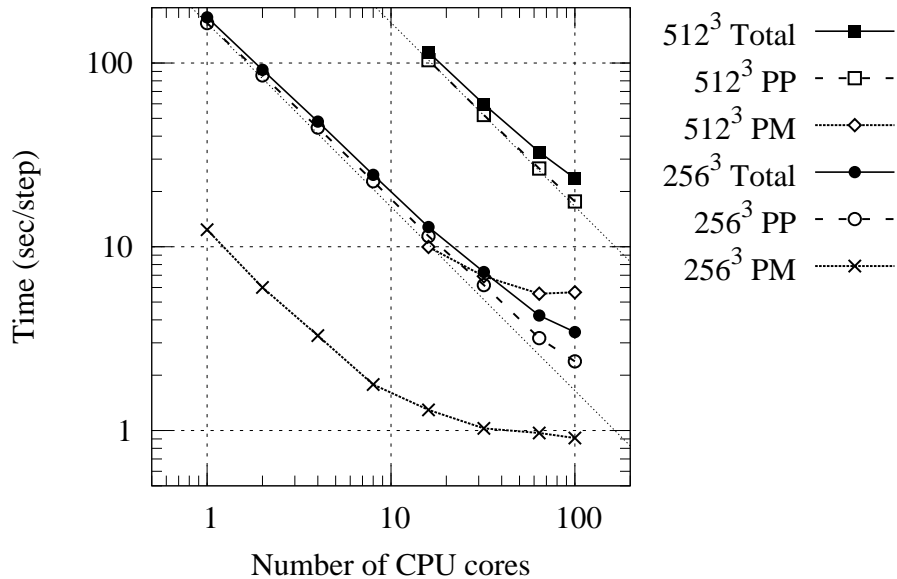}
\caption{ 
Calculation time per step  as a function of the
number of CPU cores on a PC cluster with Gigabit Ethernet.  
The curves with filled squares and circles 
show the results for total calculation time of
$512^3$, and $256^3$ dark matter simulations, respectively.
The curves with open squares and circles 
show the results for the PP part, respectively.
The curves with open argyles and crosses 
show the results for the PM part, respectively.
}
\label{np-t2}
\end{figure*}

Figure \ref{np-t2} shows the parallel speedup on a PC cluster 
of 25 nodes connected with Gigabit Ethernet.  Each node has
one Intel Core2 Quad processor (2.4GHz Q6600) and 8GB of PC6400
memory.  We can see that the speed of a PC cluster is quite similar to
that of the Cray XT4 with the same number of cores, except that the time
for the PM part is significantly longer when the number of CPU cores is
more than 32. For 100 cores, our PC cluster is about three-times
slower than the Cray XT4, while the speed of the PP part is almost the same
for all values of the number of particles and the number of processes.
This difference comes from the difference in the speed of the network.
If we use $N_{\rm PM}=N^{1/3}/4$, even with a slow Gigabit Ethernet, we
can probably use 1024 cores without seeing any significant loss of 
efficiency.

\subsubsection{Calculation Cost}

Table \ref{tab2} gives a breakdown of the calculation cost per step.
We used a run with $512^3$ dark matter particles for a measurement
of the performance. We used a snapshot at $z=0$ to
measure the CPU time for a single timestep.  The number of grid
points for the PM calculation in one dimension was $N_{\rm PM}=256$.  The
parameters of the tree parts were $\theta=0.5$, $N_{\rm crit}=300$, and
$N_{\rm leaf}=10$.  The sampling parameter was $R_{\rm samp} = 4.0
\times 10^{-4}$.  This means that $NR_{\rm samp} = 53687$ particles
were sampled for the domain update.  Here, $p$ is the number of CPU cores.

\begin{table}
\begin{center}
\caption{ Calculation time per step of our code.  }
\label{tab2}
\begin{tabular}{lrrr}
\hline \hline 
$p$ & 16 & 128 & 1024 \\ 
$N/p$ & 8388608 & 1048576 & 131072 \\ \hline 
PM (s/step) & 9.72 & 2.93 & 1.99 \\ 
\quad density assignment & 1.59 & 0.28 & 0.10 \\ 
\quad comm & 0.58 & 0.50 & 0.50 \\ 
\quad FFT & 1.12 & 1.17 & 1.17 \\ 
\quad convolution & 0.10 & 0.10 & 0.10 \\ 
\quad force interpolation & 6.33 & 0.88 & 0.12 \\ \hline 
PP (s/step) & 110.8 & 13.76 & 1.93 \\ 
\quad local tree & 3.38 & 0.38 & 0.062\\
\quad comm & 1.12 & 0.25 & 0.19 \\ 
\quad tree construction & 2.75 & 0.35 & 0.058 \\ 
\quad tree traversal & 30.42 & 3.92 & 0.44 \\ 
\quad force calculation & 73.13 & 8.86 & 1.14 \\ \hline 
Others (s/step) & 1.45 & 0.65 & 0.64 \\
\quad position update & 0.20 & 0.003 & 0.00045 \\
\quad sampling method & 0.46 & 0.08 & 0.03 \\
\quad exchange & 0.05 & 0.027 & 0.028 \\
\quad synchronization & 0.74 & 0.54 & 0.58 \\ \hline
Total (s/step) & 121.97 & 17.34 & 4.56 \\ \hline
\end{tabular}
\end{center}
\end{table}

In our current implementation, all processes have the same PM grid, and
FFT of the entire region is duplicated in all processes. Therefore,
both the communication and calculation require the time to be independent
of the number of cores. 
An obvious way to improve the performance of this part is to use
parallel FFT, such as the MPI version of the FFTW library. 
We have not implemented
this, but might consider to do so in future since the use of larger values of
$N_{\rm PM}$ allows us to use a larger opening angle, resulting in a
reduction of the CPU time for the PP part.

\subsubsection{Dependence on the Opening Angle}

\begin{figure}[t]
\centering \includegraphics[height=8cm]{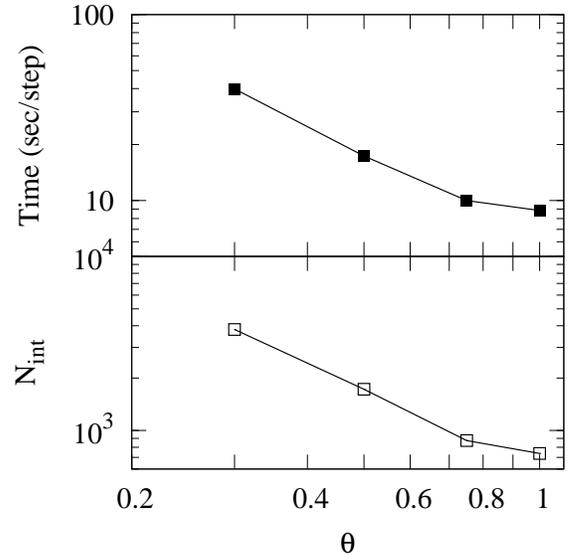}
\caption{ 
Dependence on the opening angle of the calculation time per step (top) 
and the number of force interactions per particle, $N_{\rm int}$ (bottom), 
for $512^3$ dark matter simulation.  
The number of CPU cores for the calculations is 128.
The number of force interactions per particle of the first process is plotted.
}
\label{fig:theta-t}
\end{figure}

Figure \ref{fig:theta-t} shows the CPU time per step and the average
number of force interactions, $N_{\rm int}$, per particle as functions
of the opening angle.  We used $512^3$ simulations on 32 processors with
128 CPU cores.  The parameters of the tree parts 
were $N_{\rm crit}=300$ and $N_{\rm leaf}=10$.

We can see that for small values of $\theta$, $N_{\rm int}$ is roughly
proportional to $\theta^{-2}$. Since the CPU time is dominated by the
time for the PP part, it shows almost the same behavior as $N_{\rm  int}$. 
For $\theta > 0.75$, the dependence of $N_{\rm int}$ on
$\theta$ is weak, because in this regime  $N_{\rm int}$ is
determined by $N_{\rm crit}$ and $N_{\rm leaf}$ \citep{Makino1991}.

For $\theta=0.5$, the average, minimum, and maximum of $N_{\rm int}$
are 2116, 1510, and 3156, 
respectively.  On the other hand, those of the time for the PP
part are 13.97, 13.61, 14.56, respectively.
We can see that the balance of the CPU time between teh processes is very
good, with only a 4\% difference between the average and the
maximum. This means that the loss of the performance due to non-optimal
load balancing is around 4\%.

We have not investigated the cause of this variation of the time for
the PP force, but the most likely reason is fluctuation due to
sampling, since we sample only about 400 particles per process. 

\subsubsection{Dependence on the particle distribution}

\begin{figure}[t]
\centering \includegraphics[height=9.2cm]{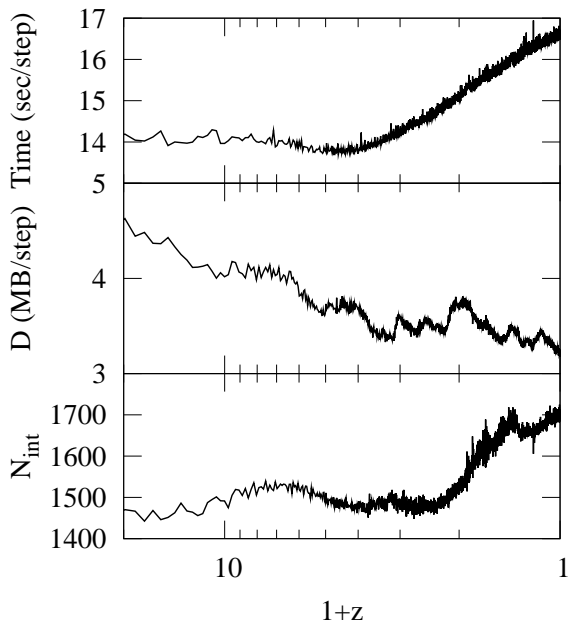}
\caption{ Dependence on the redshift of the calculation time per step (top), 
the amount of data transferred, $D$ (middle), 
and the number of force interactions per particle $N_{\rm int}$ (bottom) 
for $512^3$ dark matter simulation.  
The number of CPU cores for the calculations was 128.
The calculation time and number of force interactions per particle
and $D$ of first process are plotted.
We measured the amount of data transferred of only the PP part. 
}
\label{fig:z-t}
\end{figure}

Figure \ref{fig:z-t} shows the CPU time per step, the amount of data
transferred, $D$, and the number of force interactions, $N_{\rm int}$, as
functions of the redshift.  We used $512^3$ simulations on 32 processors
with 128 CPU cores.  Other parameters for the tree parts were
$\theta=0.5$, $N_{\rm crit}=300$, and $N_{\rm leaf}=10$.  The amount
of data transferred, $D$, is given by
\begin{equation}
D = 16 N_{\rm im} {\rm bytes},
\end{equation}
where $N_{\rm im}$ is the number of particles imported from all other
processes in the PP part. We consider only the communication for the
tree construction. 
One particle consists of the three-dimensional position and the mass, 
and the amount of data is 16 bytes per particle.
The communication for the PM grid is independent of the
particle distribution, and the amount of data transfer is 64MB per
timestep. Others are all small compared to the
tree construction. 

The calculation time is nearly constant from the start of the
calculation until $z=4$, and increases slowly afterwards as the
degree of clustering becomes higher. This increase is due to the
increase of $N_{\rm int}$.

The amount of communication, $D$, decreases as the clustering proceeds,
because of the formation of low-density voids. Since we use the tree
algorithm, the formation of high-density regions does not significantly
increase the amount of communication, while the communication need of
low-density regions is lower because of the cutoff distance of the
interaction. As a result, with TreePM, the amount of communication
decreases as the clustering proceeds. This behavior is opposite
to that of the ${\rm P^3M}$ scheme, with which the communication increases
as the clustering proceeds.

\subsubsection{Memory requirement}

The amount of required memory per particle is 48 bytes. We need to store
the position (three double precision words), the velocity (three single
precision words), a unique ID (one 64-bit integer word), and the mass (one
single precision word).  The
memory requirement per tree node is 52 bytes.  The number of nodes per
particle is $\min(1,7.5/N_{\rm leaf})$.  In addition, another 12 bytes
are required per particle for the tree-force calculation in order to
generate the morton key.  Thus the total amount of memory, $M$, required
per particle is given by the following formula:
\begin{equation}
M = 60 + 52 \cdot  \min \left(1\thinspace,\thinspace\frac{7.5}{N_{\rm leaf}} \right) {\rm bytes}.
\end{equation}
The amount of memory per PM grid point is 4.5 bytes.
It includes the mass density (one single precision word) 
and the green function table.
The green function table also needs one single precision word per table.
The number of tables is $(1/8)N_{\rm PM}^3$ owing to periodicity.
This amount is needed in all nodes.

We can use the same memory area to store the PM grid and the tree
structure, since they are not used at the same time, and both of them
are constructed from scratch at each timestep.

As discussed in subsection \ref{sec:domain}, our optimal load
balance algorithm can cause a significant imbalance in the memory
usage of up to a factor of two. However,  
as we have already shown, we can reduce the additional
amount of memory required to around 20\% or less of the total amount
of memory for particles, without any significant degradation in
performance.

\section{Discussion and Summary}

\subsection{Possible Ways to Improve Accuracy}

As we have shown in subsection 3.1, the total force error of the TreePM
method is dominated by the error of the forces from tree nodes at a 
distance of around $\eta$. Thus, one possibility of reducing the error is
to use distance-dependent opening criterion \citep{Makino1991}.
 Since the error in the limit of the uniform density
distribution is proportional to $r\theta^2 g(r/\eta)$, if we set
\begin{equation}
  \theta = \frac{\theta_0}{\sqrt{rg(r/\eta)}},
\end{equation}  
the error is evenly distributed over all nodes, and thus the error
should be minimized. It is probably necessary to set some upper-limit
value, since with the above criterion alone $\theta$ would diverge at
both ends of $r$.

A more natural approach is to include higher-order terms of
expansion. Note that the multipole expansion of force with cutoff is
different from that of a pure $1/r$ force, and should include the
spacial derivatives of the cutoff function. Thus, its implementation
differs from the usual high-order multipole moment. The second-order
moment is fairly easy to implement, and can drastically improve the
accuracy.

\subsection{Comparison with Another Code}
Here, we compare the performance of GreeM with that of GADGET-2
\citep{Springel2005}. We discuss only the scalability, because
the comparison of the absolute speed does not have much meaning if the
hardware used is different. We used the data of figure 19 in
\citet{Springel2005} for a $270^3$ dark matter simulation.  We used a
$256^3$ dark matter simulation for GreeM. This comparison is not an exact
one since the dark matter distributions and computers were different.
However it should give us some useful information.

\begin{figure}[t]
\centering \includegraphics[width=10cm]{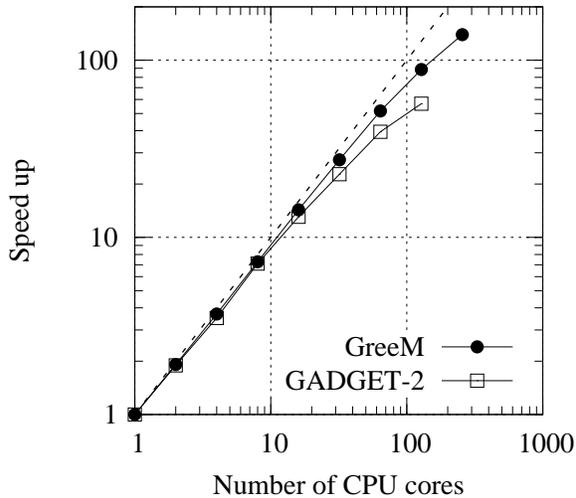}
\caption{Speed up of our code and GADGET-2 
plotted against the number of CPU cores.  
The filled circles and open squares show the result of GreeM and that of
GADGET-2, respectively
[figure 19 in \citet{Springel2005}].
}
\label{fig:diff}
\end{figure}

Figure \ref{fig:diff} shows the speed-up factors of our code and GADGET-2
as a function of the number of CPU cores. We can see that the scaling
of GreeM is better than that of GADGET-2. The most likely reason for
this difference is the difference in the load balance. Even for a very
small number of processes, the load imbalance of GADGET-2 is large (as
can be seen in their table 1). We achieved a nearly perfect load
balance, for an arbitrary number of processes.



\subsection{Summary}

In this paper, we described our new cosmological {\it N}-body
simulation code, GreeM, which uses the TreePM algorithm, and is optimized
for large parallel systems. GreeM achieves a nearly perfect load balance,
even for a very large number of cores, resulting in very good
scalability. 

GreeM runs efficiently on PC clusters, but the scalability is naturally
better on parallel computers with high-speed networks.  
The measured calculation speed on the Cray XT4 is $5 \times 10^4$ particles per second
per CPU core, if the number of particles per CPU core is larger than
$5 \times 10^5$. On a cluster of PCs with quad-core CPU and GbE
network, GreeM achieves a similar speed if the number of particles per core
is more than $3 \times 10^6$. 
Using this code, we have already performed
$1600^3$ dark matter simulation on Cray-XT4 \citep{Ishiyama2009}.  It
spent about 0.6 million CPU hours.

\bigskip
We are grateful to Kohji Yoshikawa for providing his parallel TreePM
code.  We thank Keigo Nitadori for his technical advices and providing
his Phantom GRAPE code.  T.I.  thanks Simon Portegies
Zwart and Derek Groen for helpful discussions and advices.  Numerical
computations were carried out on a Cray XT4 and a PC cluster at Center
for Computational Astrophysics, CfCA, of National Astronomical
Observatory of Japan.  T.I. is financially supported by a Research
Fellowship of the Japan Society for the Promotion of Science (JSPS)
for Young Scientists.  This research is partially supported by the
Special Coordination Fund for Promoting Science and Technology
(GRAPE-DR project), Ministry of Education, Culture, Sports, Science
and Technology, Japan.

\end{document}